\documentclass[10pt]{article}
\usepackage{multicol}
\usepackage{amsmath}
\usepackage{amsthm}
\usepackage{amsfonts}
\usepackage{graphicx}
\usepackage{float}
\usepackage{authblk}
\usepackage{geometry}
\begin{document}
\title{Microscopic study of the $^{7}$Li-nucleus potential}%
\author[1]{Wen-Di Chen}%
\author[2]{Hai-Rui Guo,\thanks{Corresponding author: guo$\_$hairui@iapcm.ac.cn}}
\author[2]{Wei-Li Sun}
\author[2]{Tao Ye}
\author[2]{Yang-Jun Ying}
\author[3]{Yin-Lu Han}
\author[3]{Qing-Biao Shen}
\affil[1]{Graduate School of China Academy of Engineering Physics, Beijing 100088, China}
\affil[2]{Institute of Applied Physics and Computational Mathematics, Beijing 100094, China}
\affil[3]{Key Laboratory of Nuclear Data, China Institute of Atomic Energy, Beijing 102413, China}
\maketitle

\textbf{Abstract}  The optical potential without any free parameters for $^7$Li-nucleus interaction system is studied in a microscopic approach. It is obtained by folding the microscopic optical potentials of the constituent nucleons of $^{7}$Li over their density distributions. We employ an isospin-dependent nucleon microscopic optical potential, which is based on the Skyrme nucleon-nucleon effective interaction and derived by using the Green's function method, to be the nucleon optical potential. Harmonic oscillator shell model is used to describe the internal wave function of $^7$Li and get the nucleon density distribution. The $^7$Li microscopic optical potential is used to predict the reaction cross sections and elastic scattering angular distributions for target range from $^{27}$Al to $^{208}$Pb and energy range below 450 MeV. Generally the results can reproduce the measured data reasonably well. In addition, the microscopic optical potential is comparable to a global phenomenological optical potential in fitting the presently existing measured data generally.



This manuscript is to be published in Chinese Physics C.

\section{Introduction}

Optical potential is an usual and basic tool used in the dynamic analyses of nuclear reactions. Nowadays, most of the optical potentials are phenomenological. They have some parameters, and are determined by fitting experimental data. When the experimental data are not sufficient, it is difficult to get reliable phenomenological optical potential. In contrast, the microscopic optical potential (MOP) is derived from nucleon-nucleon interaction theoretically, has no free parameters, and does not rely on the experimental data. Therefore, to obtain optical potentials in microscopic approach is a goal of the nuclear physics. It is of great significance for the analyses of nuclear reactions lacking experimental data.

The studies of nuclear reactions involving light-particle projectile or ejectile are an important part of nuclear physics and very useful for practical applications. Thus, we have already obtained the MOPs for nucleon \cite{Shen2009}, deuteron \cite{Guo2010a}, triton \cite{Guo2014b}, and $^{3,4,6}$He \cite{Guo2009c,Guo2011d,Guo2017e}. Recent years, the weakly bound $^{7}$Li induced reactions has been a subject of great interest. Breakup, complete and incomplete fusion, and some other reaction mechanisms are concerned by the experimental and theoretical nuclear physicists \cite{Parkar2018, Gautam2017}. The $^{7}$Li optical potential is required in the theoretical analyses.

Up to now, there are some $^{7}$Li optical potentials to analyze the experimental data. A semi-microscopic optical potential, whose real part is generated by double folding model and nucleon-nucleon effective interaction and imaginary part is in the Woods-Saxon form, is given by Woods et al. \cite{Woods1982} and used to analyze the elastic scattering data for $^{15}$N and $^{25}$Mg target. Deshmukh et al. \cite{Deshmukh2011} provided a Wood-Saxon form optical potential, while it can only be used for $^{116}$Sn. An optical potential provided by Camacho et al. \cite{GomezCamacho2010} meets the dispersion relation of real part and imaginary part, but it is only suitable for $^{28}$Si target. Recently, Xu et al. \cite{Yongli2018} provided a new global phenomenological optical potential (GOP) based on the presently existing experimental data, which is applicable to a more extensive incident energy and target region.

Since the measured $^{7}$Li scattering data are not sufficient up to now, a $^{7}$Li MOP is obtained in the present work by folding the MOPs of its internal nucleons over their density distributions. The isospin-dependent nonrealistic nucleon MOP derived by using the Green's function method in our previous work \cite{Shen2009,Xu2014,Xu2017,Xu2017a} is adopted to be the MOP for the constituent nucleons. Shell model is applied to construct the internal wave function and generate the nucleon density distributions. The $^{7}$Li elastic-scattering angular distributions and reaction cross sections are calculated by the MOP and compared with the experimental data and the results calculated by the GOP \cite{Yongli2018}.

This paper is organized as follow: the theoretical model and formulas of the MOP are presented in Sec. 2; the calculated results and analysis are provided in Sec. 3; the summary and conclusion are given in Sec. 4 finally.

\section{Theoretical model}
The MOP for $^{7}$Li is generated by the folding model \cite{Satchler1979} and expressed as
\begin{equation}\label{e-fold}
  U(\vec R) = \int {{U_n}(\vec R + \vec r){\rho _n}(\vec r) + {U_p}(\vec R + \vec r){\rho _p}(\vec r)d\vec r},
\end{equation}
where
\begin{equation}\label{e-rou}
  \int {{\rho _n}d\vec r}  = N;\int {{\rho _p}d\vec r}  = Z.
\end{equation}
$U_n$ and $U_p$ represent the MOPs for neutron and proton respectively. $\rho _n$ and $\rho _p$ are the density distributions of neutron and proton in the ground state $^{7}$Li respectively. $\vec R$ is the relative coordinate between the centers of mass of the target and $^{7}$Li, and $\vec r$ is the internal coordinate of $^7$Li.

The isospin-dependent nonrealistic nucleon MOP \cite{Shen2009,Xu2014,Xu2017,Xu2017a} is adopted to be $U_n$ and $U_p$ and here is a brief introduction for it. From the perspective of many-body theory, the nucleon optical potential is equivalent to the mass operator of the single-particle Green's function \cite{Bell1959}. Based on the Skyrme nucleon-nucleon effective interaction SKC16 \cite{Xu2017}, which is able to describe the nuclear matter properties, ground state properties and neutron-nucleus scattering well simultaneously, the first- and second-order mass operators of single-particle Green's function were derived through the nuclear matter approximation and the local density approximation. The real part of the nucleon MOP was denoted by the first-order mass operator and the imaginary part of the nucleon MOP was denoted by the imaginary part of the second-order mass operator. The incident energy of nucleon is regarded as one seventh of the incident energy of $^7$Li.

Shell model is adopted to give an appropriate nucleon density in $^{7}$Li. Since a 1p-shell model space can well describe its structure \cite{Gu1996} and we only concern the ground-state properties of $^{7}$Li here, harmonic oscillator potential is adopted to describe the mean interaction for the nucleons in $^7$Li, and the internal Hamiltonian of $^7$Li is expressed as
\begin{equation}\label{e-ham}
  {H_{{}^7Li}} = \sum\limits_{i = 1}^7 {{T_i}}  + \sum\limits_{i = 1}^7 {\frac{1}{2}m{\omega ^2}r_i^2} ,
\end{equation}
where $m$ is the nucleon mass and $r_i$ is coordinate of the $i$th nucleon in $^{7}$Li relative to the center of mass of $^{7}$Li. $T_i$ represents the kinetic energy of the $i$th nucleon.

As the harmonic oscillator potential is used in the shell model, the ground-state wave function of $^{7}$Li is expressed as
\begin{equation}\label{e-phigs}
{\Phi _{g.s.}} = N \mathcal{A} \left\{ {({{\vec r}_6}\cdot{{\vec r}_7}){r_5}Y_1^{\mu}({\hat r}_5)\exp \{  - \frac{\beta }{2}\sum\limits_{i = 1}^7 {r_i^2} \} \zeta } \right\},
\end{equation}
where $\mathcal{A}$ is the antisymmetrization operator of the nucleons and $N$ is the normalization factor. $\zeta $ represents the spin and isospin part. ${\Phi _{g.s.}}$ is determined by the parameter $\beta = m \omega / \hbar $, under the conditions of meeting antisymmetrization, spin and parity ($I^{\pi}=3/2^{-}$). On base of the constraint condition
\begin{equation}\label{e-constraint}
\sum\limits_{i = 1}^7 {{{\vec r}_i}}  = 0,
\end{equation}
a set of Jacobi coordinates is used to replace $r_i$ and expressed as
\begin{eqnarray}\label{e-jacobi}
&& {{\vec r}_1} = \frac{1}{2}{{\vec \xi }_1} + \frac{1}{3}{{\vec \xi }_2} + \frac{1}{4}{{\vec \xi }_3} + \frac{1}{3}{{\vec \xi }_5} + \frac{1}{7}{{\vec \xi }_6}, \nonumber\\[1mm]
&& {{\vec r}_2} =  - \frac{1}{2}{{\vec \xi }_1} + \frac{1}{3}{{\vec \xi }_2} + \frac{1}{4}{{\vec \xi }_3} + \frac{1}{3}{{\vec \xi }_5} + \frac{1}{7}{{\vec \xi }_6}, \nonumber\\[1mm]
&& {{\vec r}_3} =  - \frac{2}{3}{{\vec \xi }_2} + \frac{1}{4}{{\vec \xi }_3} + \frac{1}{3}{{\vec \xi }_5} + \frac{1}{7}{{\vec \xi }_6}, \nonumber\\[1mm]
&& {{\vec r}_4} =  - \frac{3}{4}{{\vec \xi }_3} + \frac{1}{3}{{\vec \xi }_5} + \frac{1}{7}{{\vec \xi }_6}, \nonumber\\[1mm]
&& {{\vec r}_5} =  - \frac{6}{7}{{\vec \xi }_6},  \nonumber\\[1mm]
&& {{\vec r}_6} = \frac{1}{2}{{\vec \xi }_4} - \frac{2}{3}{{\vec \xi }_5} + \frac{1}{7}{{\vec \xi }_6},  \nonumber\\[1mm]
&& {{\vec r}_7} =  - \frac{1}{2}{{\vec \xi }_4} - \frac{2}{3}{{\vec \xi }_5} + \frac{1}{7}{{\vec \xi }_6}.
\end{eqnarray}

The value of $\beta $ is determined by
\begin{equation}\label{e-rms}
\left\langle {r_{rms}^2} \right\rangle  = \left\langle {{\Phi _{g.s.}}} \right|{\frac{1}{7}\sum\limits_{i = 1}^7 {r_i^2}}{\left| {{\Phi _{g.s.}}} \right\rangle } ,
\end{equation}
where $\sqrt {\left\langle {r_{rms}^2} \right\rangle } $ is the nuclear matter root-mean-square radius of $^{7}$Li and set as 2.50 fm which was obtained by fitting the reaction cross section in Ref. \cite{Tanihata1985}. It will be convenient to rewrite the Eq. (\ref{e-phigs}) as below
\begin{equation}\label{e-rewrite}
{\Phi _{g.s.}} = N\mathcal{A}\left\{ {{\phi _1}(1234){\phi _2}(5){\phi _3}(67)\zeta } \right\},
\end{equation}
where
\begin{eqnarray}\label{e-phi123}
&& {\phi _1}(1234) = \exp \{  - \frac{\beta }{2}\sum\limits_{i = 1}^4 {r_i^2} \}, \nonumber\\[1mm]
&& {\phi _2}(5) = {r_5}Y_1^m({{\hat r}_5})\exp \{  - \frac{\beta }{2}r_5^2\}, \nonumber\\[1mm]
&& {\phi _3}(67) = ({{\vec r}_6}\cdot{{\vec r}_7})\exp \{  - \frac{\beta }{2}(r_6^2 + r_7^2)\}.
\end{eqnarray}
Then we can get a detailed expression for $\left\langle {r_{rms}^2} \right\rangle$,
\begin{equation}\label{e-rms-2}
\left\langle {r_{rms}^2} \right\rangle  = \frac{{{A_0} - 2{A_1} - {A_2} + 2{A_3} + {A_4} - {A_5}}}{{{N_0} - 2{N_1} - {N_2} + 2{N_3} + {N_4} - {N_5}}},
\end{equation}
where
\begin{eqnarray}\label{e-A1-A5}
&& {A_0} = \left\langle {{\phi _1}(1234){\phi _2}(5){\phi _3}(67)} \right|{{\hat O}_A}{\left| {{\phi _1}(1234){\phi _2}(5){\phi _3}(67)} \right\rangle _\xi }, \nonumber\\[1mm]
&& {A_1} = \left\langle {{\phi _1}(1234){\phi _2}(5){\phi _3}(67)} \right|{{\hat O}_A}{\left| {{\phi _1}(1264){\phi _2}(5){\phi _3}(37)} \right\rangle _\xi }, \nonumber\\[1mm]
&& {A_2} = \left\langle {{\phi _1}(1234){\phi _2}(5){\phi _3}(67)} \right|{{\hat O}_A}{\left| {{\phi _1}(5234){\phi _2}(1){\phi _3}(67)} \right\rangle _\xi }, \nonumber\\[1mm]
&& {A_3} = \left\langle {{\phi _1}(1234){\phi _2}(5){\phi _3}(67)} \right|{{\hat O}_A}{\left| {{\phi _1}(5264){\phi _2}(1){\phi _3}(37)} \right\rangle _\xi }, \nonumber\\[1mm]
&& {A_4} = \left\langle {{\phi _1}(1234){\phi _2}(5){\phi _3}(67)} \right|{{\hat O}_A}{\left| {{\phi _1}(1267){\phi _2}(5){\phi _3}(34)} \right\rangle _\xi }, \nonumber\\[1mm]
&& {A_5} = \left\langle {{\phi _1}(1234){\phi _2}(5){\phi _3}(67)} \right|{{\hat O}_A}{\left| {{\phi _1}(5267){\phi _2}(1){\phi _3}(34)} \right\rangle _\xi }, \nonumber\\[1mm]
&& {{\hat O}_A} = \frac{1}{7}\sum\limits_{i = 1}^7 {r_i^2}.
\end{eqnarray}
${\left\langle {...} \right\rangle _\xi }$ means that the coordinates $\{ \vec r_i, i=1-7 \}$ are replaced by the Jacobi coordinates $\{ \vec \xi _i, i=1-6 \}$. The formulas for $N_i$ are the same as $A_i$ while ${{\hat O}_A}$ is replaced by ${{\hat O}_N}=1$. Thus we can get that
\begin{equation}\label{e-rms-beta}
\left\langle {r_{rms}^2} \right\rangle  = \frac{{12}}{{7\beta }}
\end{equation}
and therefore $\beta$=0.2743 fm${^{ - 2}}$.

\begin{figure}[htbp]
  \centering
  \includegraphics[width=0.6\columnwidth]{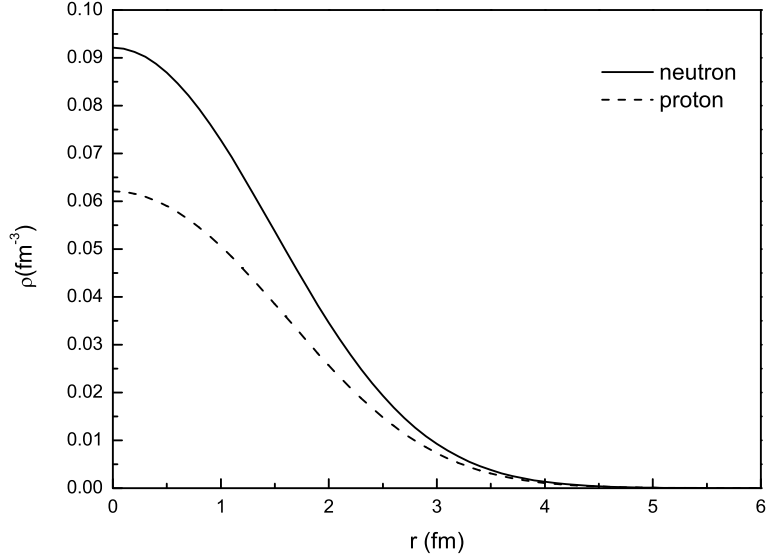}
  \caption{Neutron ($\rho _n$) and proton ($\rho _p$) density distributions.}
  \label{f-den}
\end{figure}

$\rho _n$ and $\rho _p$ are defined as
\begin{equation}\label{e-rou-def}
{\rho _{n(p)}}(\vec r) = \left\langle {{\Phi _{g.s.}}} \right| {\sum\limits_{i = 1}^7 {\delta (\vec r - {{\vec r}_i}){\delta _{{\tau _{n(p)}},{\tau _i}}}} } {\left| {{\Phi _{g.s.}}} \right\rangle },
\end{equation}
where $\tau _i$ is the isospin of $i$th nucleon. $\tau _n$ and $\tau _p$ are the isospin of neutron and proton respectively. It would be convenient to calculate $\rho _i$ firstly, whose formula is the same as Eq. (\ref{e-rms-2}) while only ${{\hat O}_A}$ is replaced by ${{\hat O}_{\rho ,i}}={\delta (\vec r - {{\vec r}_i})}$. $\rho _p = \rho _1 + \rho _2 + \rho _5$, $\rho _n = \rho _3 + \rho _4 + \rho _6 + \rho _7$, and they have analytical expressions as
\begin{equation}\label{e-rou-form}
{\rho _{n(p)}}(\vec r) = \left( {{a_{n(p)}} + {b_{n(p)}}{r^2}} \right)\exp \left(  - \frac{7}{{6}}\beta {r^2}\right),
\end{equation}
where $a_n$=0.0921 fm$^{-3}$, $a_p$=0.0621 fm$^{-3}$, $b_n$=0.0081 fm$^{-5}$ and $b_p$=0.0076 fm$^{-5}$. The density distributions are plotted in Fig. \ref{f-den}.

\section{Calculated result and analysis}
\begin{figure}[htbp]
  \centering
  \includegraphics[width=0.6\columnwidth]{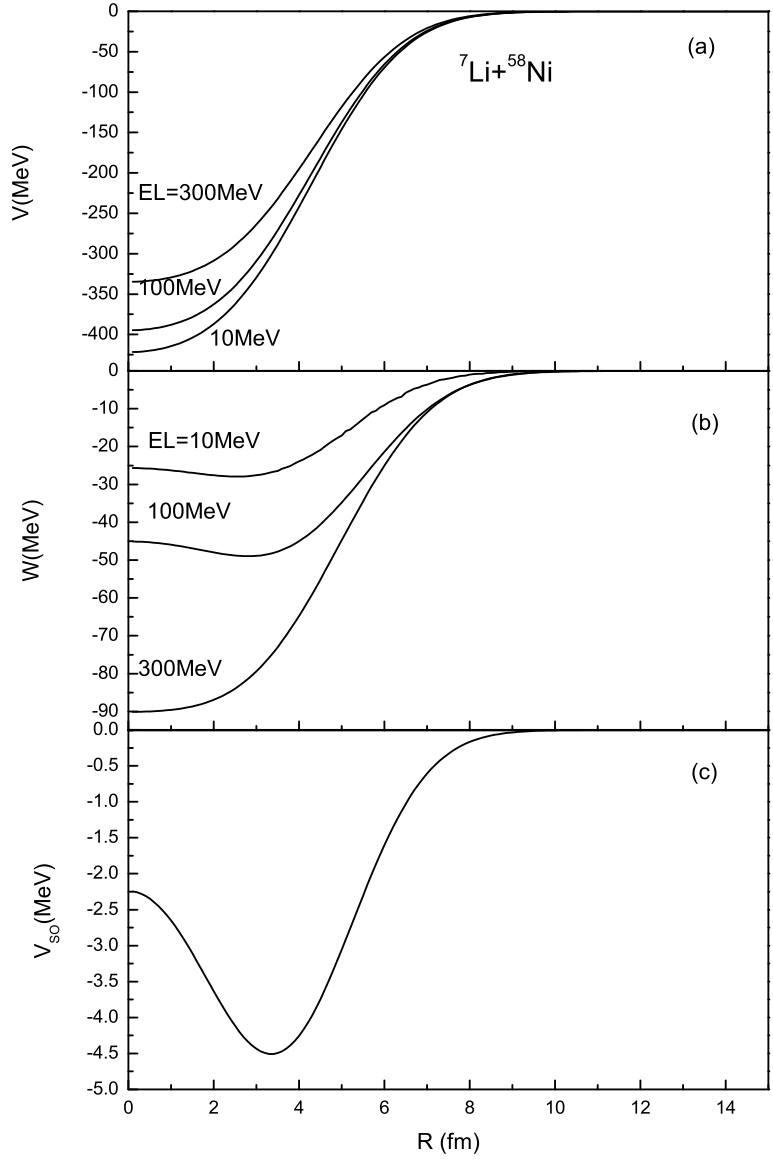}
  \caption{The MOP for $^7$Li+$^{58}$Ni system (a) the real part ($V$), (b) the imaginary part ($W$) and (c) the real part of the spin-orbit potential ($V_{so}$).}
  \label{f-pot}
\end{figure}
The MOP for $^7$Li+$^{58}$Ni collision system at incident $^{7}$Li energies of 10 MeV, 100 MeV and 300 MeV is shown in Fig. \ref{f-pot} as an example. The depth of the real part ($V$) decreases with the increase of the radius and energy. However the depth of the imaginary part ($W$) increases a little first and decreases as the radius increases at EL=10, 100 MeV, while it decreases monotonously with the increase of the radius at a higher incident energy, 300 MeV. That means the contribution of $W$ changes from the dominant surface absorption to the volume absorption as the incident energy increases. The real part of the spin-orbit potential $V_{so}{\vec s} \cdot {\vec l}$ is also obtained by folding model and $V_{so}$ is shown in Fig. \ref{f-pot}, while the imaginary part of the spin-orbit potential is omitted as it is usually very small.

The $^{7}$Li elastic-scattering angular distributions and reaction cross sections are predicted by using the MOP. Comparisons with experimental data and the results calculated by the GOP \cite{Yongli2018} are made.

\begin{figure}[htbp]
  \centering
  \includegraphics[width=0.6\columnwidth]{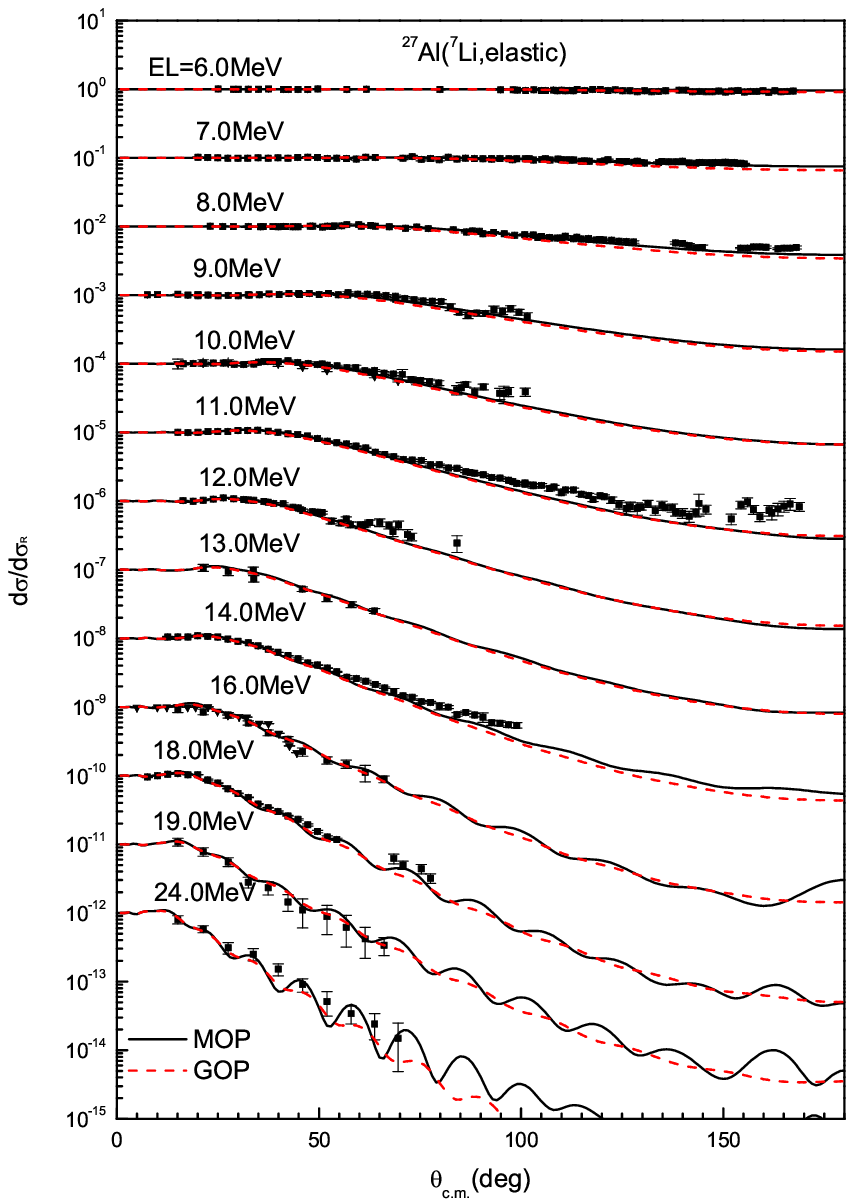}
  \caption{(color online) Calculated elastic-scattering angular distributions in the Rutherford ratio for $^{27}$Al compared with experimental data \cite{Figueira2006,Kalita2006}. The solid and dash lines denote the results calculated by the MOP and the GOP \cite{Yongli2018} respectively. The results from top to bottom are multiplied respectively by 10$^{0}$, 10$^{-1}$, 10$^{-2}$...}
  \label{f-Al27}
\end{figure}
Fig. \ref{f-Al27} shows the elastic-scattering angular distribution for $^{27}$Al target at incident energies from 6.0 MeV to 24.0 MeV. The result calculated by the MOP is in good agreement with experimental data \cite{Figueira2006,Kalita2006} except for the underestimation at EL=11.0 MeV for large angles.  In addition, the MOP result fits the experimental data a little better than the result calculated by the GOP \cite{Yongli2018} below 14 MeV at larger angles.

\begin{figure}[htbp]
  \centering
  \includegraphics[width=0.6\columnwidth]{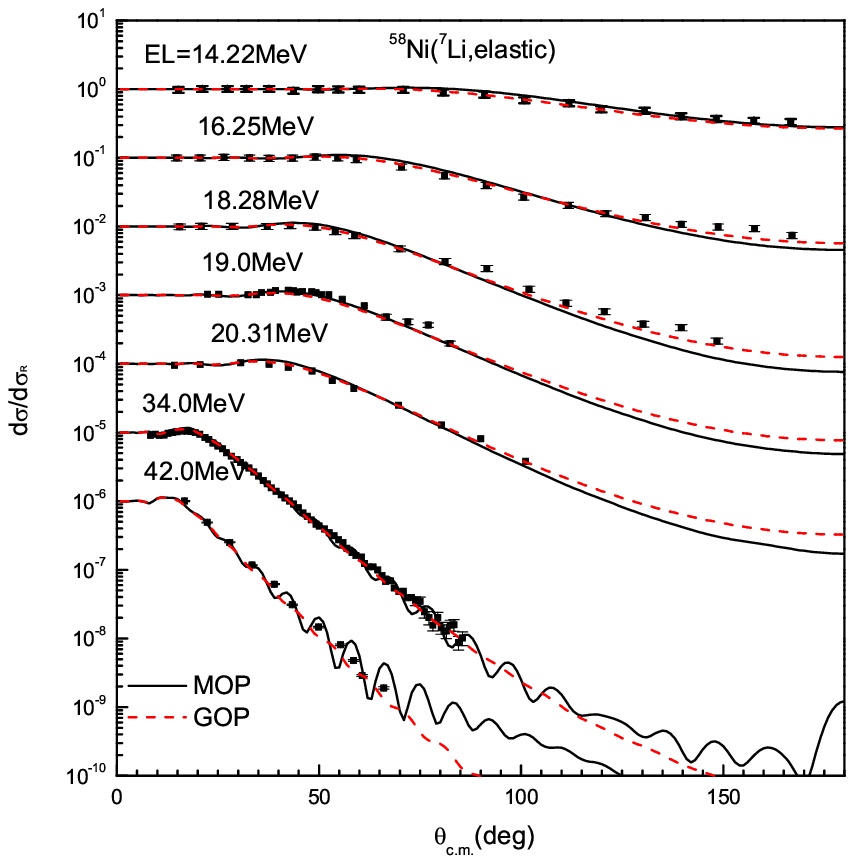}
  \caption{(color online) Same as Fig. \ref{f-Al27} but for $^{58}$Ni. The experimental data are taken from Refs.  \cite{Glover1980,Andronic1999,Zerva2012}. The results from top to bottom are multiplied respectively by 10$^{0}$, 10$^{-1}$, 10$^{-2}$...}
  \label{f-Ni58}
\end{figure}
The calculated elastic-scattering angular distribution for $^{58}$Ni target at incident energies from 14.22 MeV to 42.0 MeV is plotted in Fig. \ref{f-Ni58}. The MOP reproduces the experimental data \cite{Glover1980,Andronic1999,Zerva2012} well except the slight underestimation above 70 degrees at 16.25 and 18.28 MeV, where the GOP performs a little better.

\begin{figure}[htbp]
  \centering
  \includegraphics[width=0.6\columnwidth]{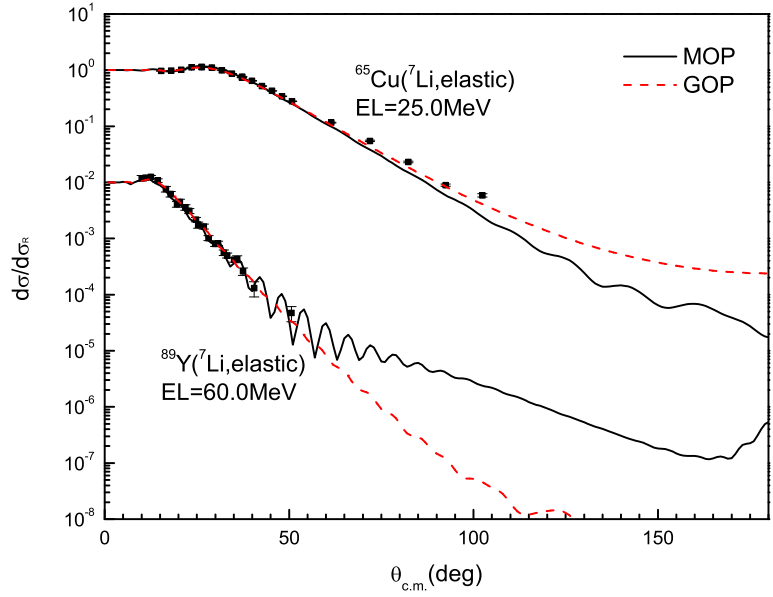}
  \caption{(color online) Same as Fig. \ref{f-Al27} but for $^{65}$Cu and $^{89}$Y. The experimental data are taken from Refs. \cite{Shrivastava2006,Wadsworth1983}. The data for $^{89}$Y are multiplied by 10$^{-2}$.}
  \label{f-Cu65-Y89}
\end{figure}
The elastic-scattering angular distributions  for $^{65}$Cu at incident energy 25.0 MeV and $^{89}$Y at incident energy 60.0 MeV are shown in Fig. \ref{f-Cu65-Y89}. It can be observed for $^{65}$Cu that the theoretical result from the MOP is lower than the measured values \cite{Shrivastava2006} above 70 degrees. Reasonable agreement with the experimental data \cite{Wadsworth1983} on $^{89}$Y is obtained.

\begin{figure}[htbp]
  \centering
  \includegraphics[width=0.6\columnwidth]{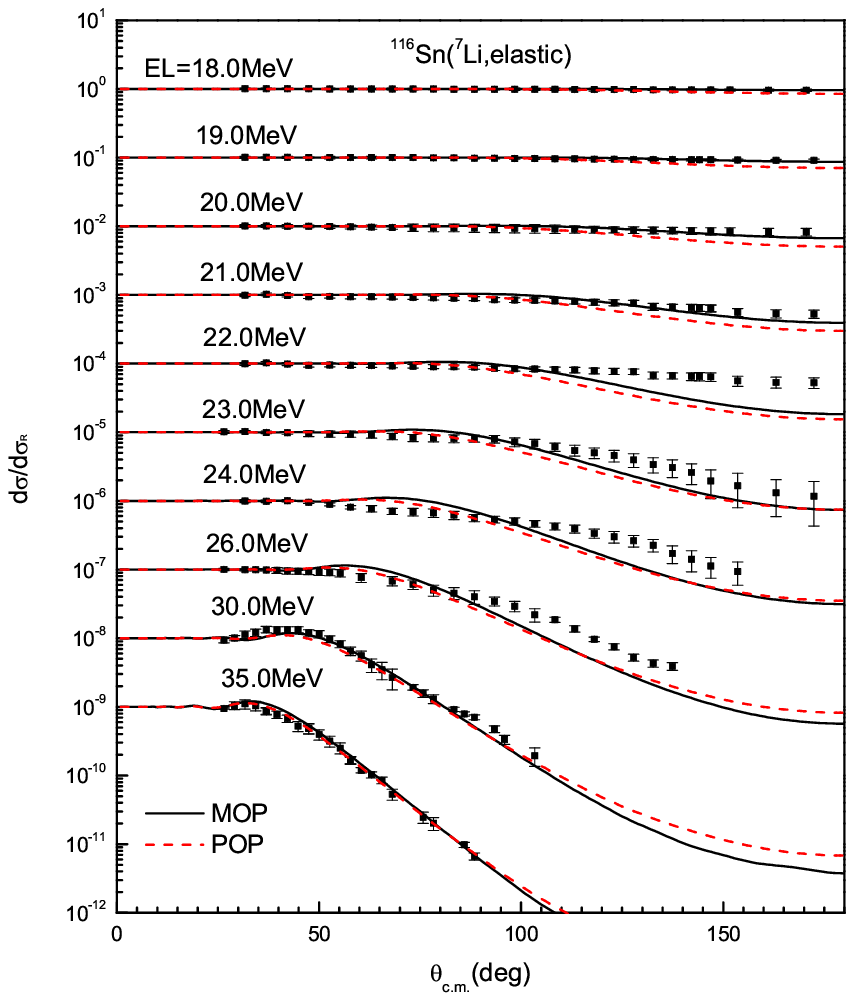}
  \caption{(color online) Same as Fig. \ref{f-Al27} but for $^{116}$Sn. The experimental data are taken from Ref.  \cite{Deshmukh2011}. The results from top to bottom are multiplied respectively by 10$^{0}$, 10$^{-1}$, 10$^{-2}$...}
  \label{f-Sn116}
\end{figure}
Fig. \ref{f-Sn116} shows the elastic-scattering angular distribution for $^{116}$Sn target at incident energies from 18.0 MeV to 35.0 MeV. The calculated result from the MOP is in good agreement with experimental data \cite{Deshmukh2011} except for those at incident energies 22.0 MeV, 24.0 MeV and 26.0 MeV in large angles. It can be seen that the MOP reproduces the measurements a little better than the GOP does at relatively lower energies.

\begin{figure}[htbp]
  \centering
  \includegraphics[width=0.6\columnwidth]{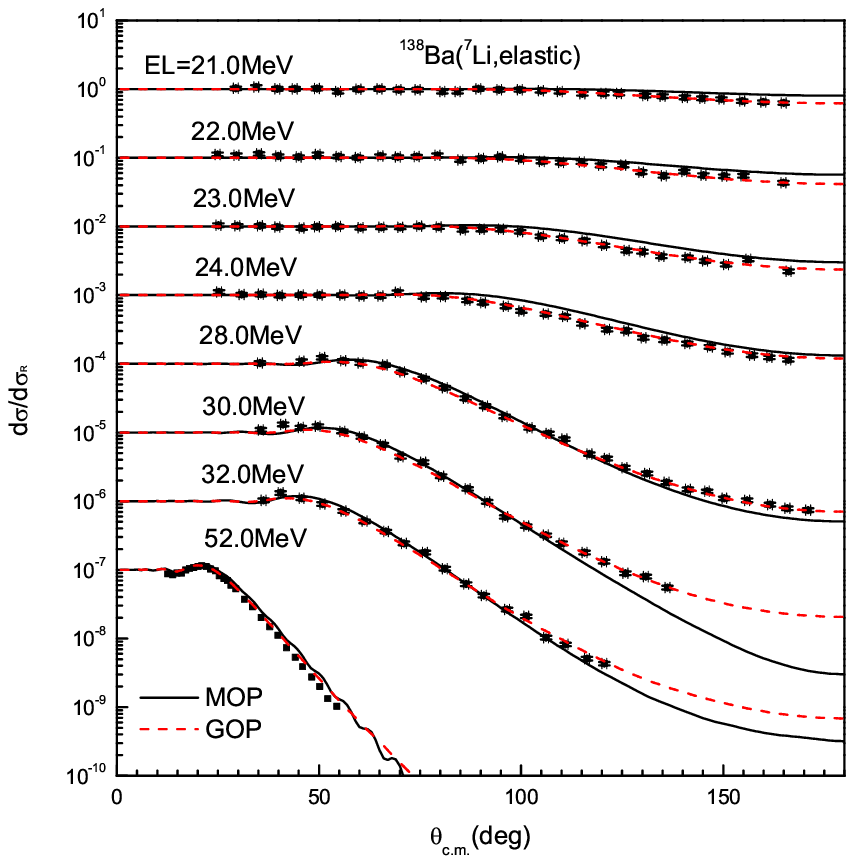}
  \caption{(color online) Same as Fig. \ref{f-Al27} but for $^{138}$Ba. The experimental data are taken from Refs.  \cite{Clark1980,Maciel1999}. The results from top to bottom are multiplied respectively by 10$^{0}$, 10$^{-1}$, 10$^{-2}$...}
  \label{f-Ba138}
\end{figure}
The calculated elastic-scattering angular distribution for $^{138}$Ba targets is compared with experimental data \cite{Clark1980,Maciel1999} in Fig. \ref{f-Ba138}. When the scattering angles are less than 80 degrees, good agreement with experimental data is obtained for the MOP. The GOP works better at larger angles.

\begin{figure}[htbp]
  \centering
  \includegraphics[width=0.6\columnwidth]{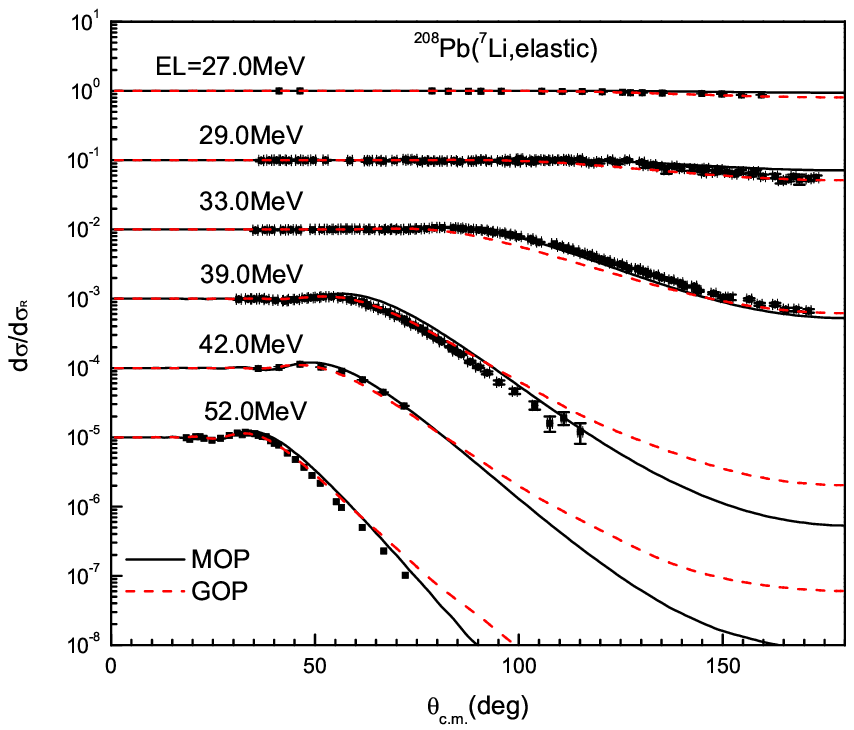}
  \caption{(color online) Same as Fig. \ref{f-Al27} but for $^{208}$Pb. The experimental data are taken from Refs.  \cite{Parkar2008,Keeley1994,Gupta2001,Zeller1979}. The results from top to bottom are multiplied respectively by 10$^{0}$, 10$^{-1}$, 10$^{-2}$...}
  \label{f-Pb208}
\end{figure}
In Fig. \ref{f-Pb208}, the calculated elastic-scattering angular distribution for $^{208}$Pb is shown from 27.0 MeV to 52.0 MeV. The MOP result is in satisfying agreement with experimental data \cite{Parkar2008,Keeley1994,Gupta2001,Zeller1979} and comparable to the GOP result in fitting the measured data, except for the case at 39 MeV above 70 degrees.

\begin{figure}[htbp]
  \centering
  \includegraphics[width=0.6\columnwidth]{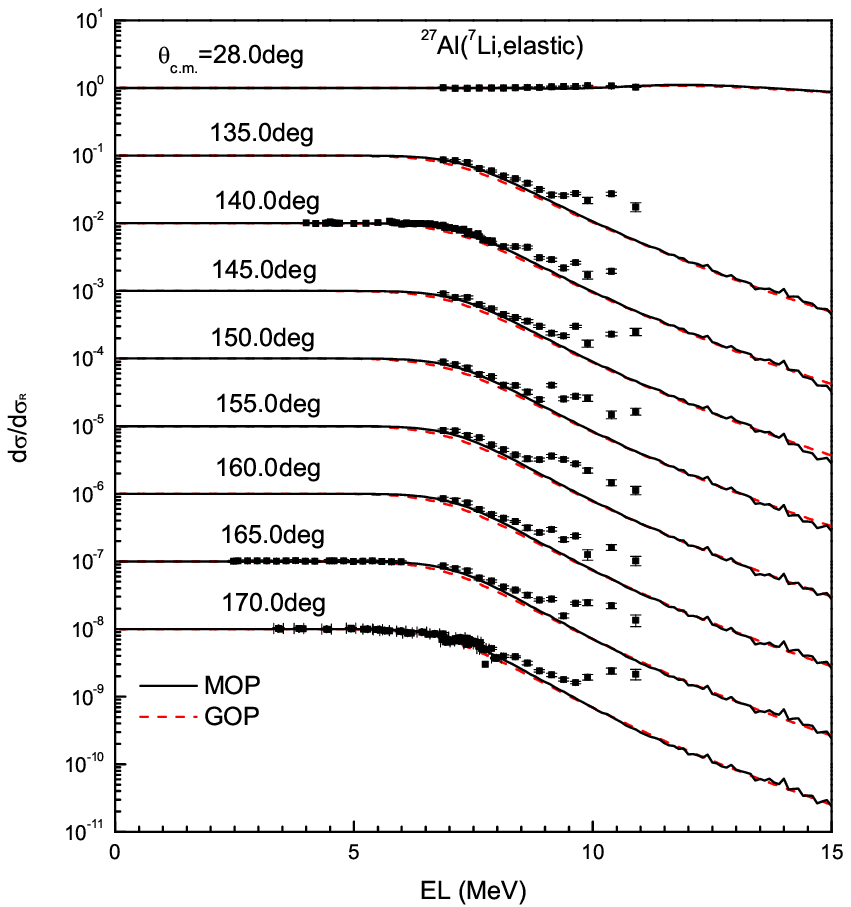}
  \caption{(color online) Calculated elastic-scattering angular distributions in the Rutherford ratio for $^{27}$Al at some scattering angles compared with experimental data \cite{Raisanen1993a,Rauhala1994a,Nurmela1999,Mayer2003,Abriola2010}. The solid and dash lines denote the results calculated by the MOP and the GOP \cite{Yongli2018} respectively. The results from top to bottom are multiplied respectively by 10$^{0}$, 10$^{-1}$, 10$^{-2}$...}
  \label{f-sa-Al27}
\end{figure}
Fig. \ref{f-sa-Al27} shows the elastic-scattering angular distribution at some specific scattering angles for $^{27}$Al target. The calculated result by the MOP is slightly larger than that by the GOP at incident energies below 15 MeV and has a little better agreement with the measured values \cite{Raisanen1993a,Rauhala1994a,Nurmela1999,Mayer2003} when EL$\leq $9 MeV. However all calculated results from the MOP and the GOP underestimate the experiment data \cite{Abriola2010} at incident energies from 9 to 11 MeV for large angles.

\begin{figure}[htbp]
  \centering
  \includegraphics[width=0.6\columnwidth]{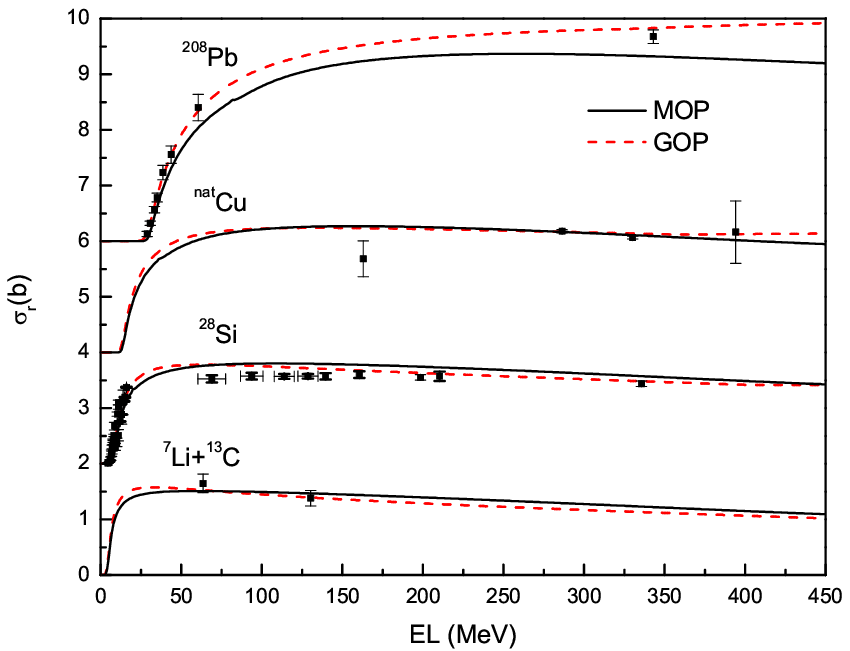}
  \caption{(color online) Reaction cross sections calculated by the MOP compared with experimental data for $^{13}$C \cite{Carstoiu2004}, $^{28}$Si \cite{Warner1996,Kashihira2005,Pakou2007,Pakou2009,Musumarra2010}, $^{nat}$Cu \cite{Saint-Laurent1989} and $^{208}$Pb \cite{Keeley1994,Figueira2010} and the results calculated by the GOP \cite{Yongli2018}. The data are shifted upwards by adding 0, 2, 4 and 6 b respectively. The solid and dash lines denote the results calculated by the MOP and the GOP respectively.}
  \label{f-s-450}
\end{figure}
\begin{figure}[htbp]
  \centering
  \includegraphics[width=0.6\columnwidth]{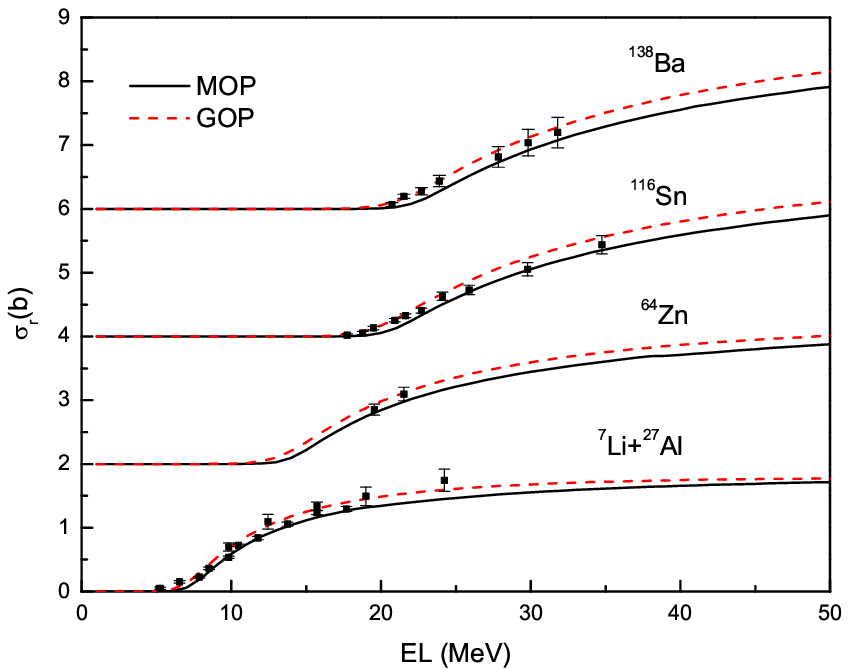}
  \caption{(color online) Same as Fig. \ref{f-s-450} but for $^{27}$Al, $^{64}$Zn, $^{116}$Sn and $^{138}$Ba. The experimental data are taken from Refs. \cite{Kalita2006,Benjamim2007,Gomes2005,Deshmukh2011,Maciel1999}.}
  \label{f-s-50}
\end{figure}
The reaction cross sections of $^7$Li induced reactions on $^{13}$C, $^{27}$Al, $^{28}$Si, $^{64}$Zn, $^{nat}$Cu, $^{116}$Sn, $^{138}$Ba, $^{208}$Pb are also calculated and shown in Fig. \ref{f-s-450} and Fig. \ref{f-s-50}. Fig. \ref{f-s-450} presents the results for $^{13}$C, $^{28}$Si , $^{nat}$Cu and $^{208}$Pb. The theoretical result for $^{13}$C is within the measurement error range \cite{Carstoiu2004}. The MOP result for $^{28}$Si is in good agreement with experimental data \cite{Warner1996,Kashihira2005,Pakou2007,Pakou2009,Musumarra2010} below 30 MeV but becomes a little larger from 90 MeV to 200 MeV. The reaction cross section for $^{nat}$Cu is obtained by averaging the reaction cross sections for $^{63}$Cu and $^{65}$Cu over the natural abundance. It can be seen that the MOP result is in good agreement with experimental data \cite{Saint-Laurent1989} except for the energy point of 160 MeV. The MOP result for $^{208}$Pb reproduces the experimental data \cite{Keeley1994,Figueira2010} reasonably below 70 MeV but gives an underestimation at 300 MeV. In Fig. \ref{f-s-50}, it can be observed that the MOP reproduces the experimental data for $^{27}$Al \cite{Kalita2006,Benjamim2007}, $^{64}$Zn \cite{Gomes2005}, $^{116}$Sn \cite{Deshmukh2011}, and $^{138}$Ba \cite{Maciel1999} well. The MOP results are comparable to the GOP results in fitting the measured reaction cross sections except for $^{208}$Pb.

\begin{figure}[htbp]
  \centering
  \includegraphics[width=0.6\columnwidth]{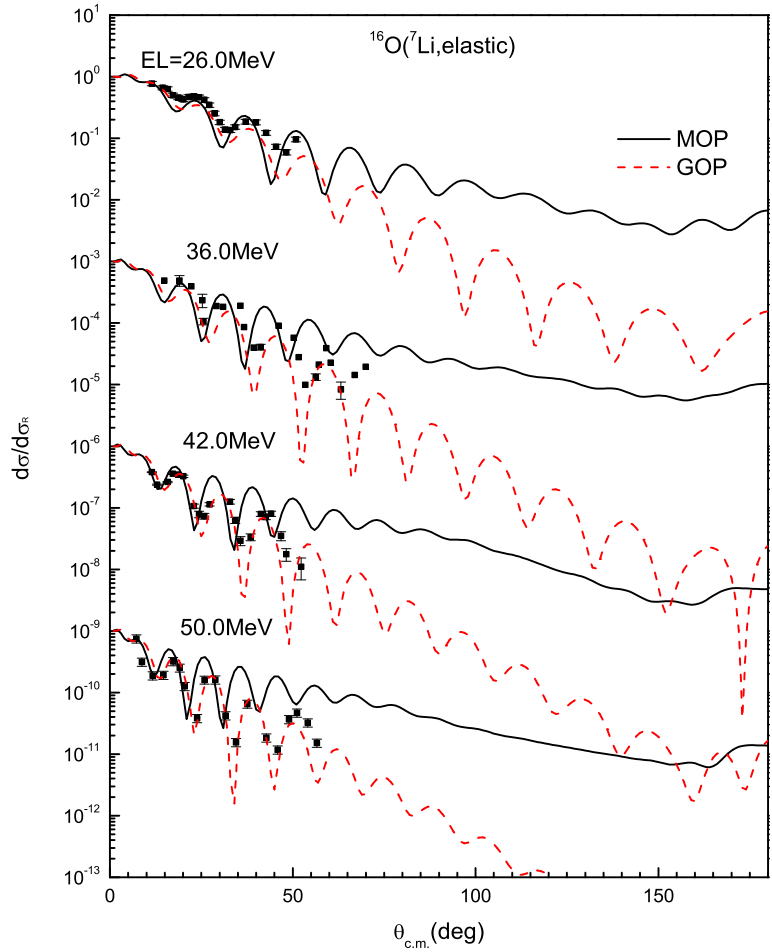}
  \caption{(color online) Same as Fig. \ref{f-Al27} but for $^{16}$O. The experimental data are taken from Refs. \cite{Guo2012,Schumacher1973,Rudchik2007,Cook1984}. The results from top to bottom are multiplied respectively by 10$^{0}$, 10$^{-3}$, 10$^{-6}$ and 10$^{-9}$.}
  \label{f-O16}
\end{figure}
The application of the MOP to the prediction of $^{7}$Li elastic scattering from light target nuclei is also tried. The elastic-scattering angular distribution for $^{16}$O is calculated and compared with experimental data \cite{Guo2012,Schumacher1973,Rudchik2007,Cook1984} as shown in Fig. \ref{f-O16}. The theoretical result from the MOP is only consistent with the magnitude of measured data in forward angles and gives an overestimation in relative larger angles. Therefore, the MOP is not suitable for the light nuclei. On the one hand, it may be interpreted that the Negele's nuclear density \cite{Negele1970} adopted to calculate the nucleon MOP \cite{Shen2009,Xu2014,Xu2017,Xu2017a} is not suitable for light nuclei. On the other hand, light nucleus, such as $^{16}$O, has its unique structure characteristics and reaction mechanism \cite{Sun2008,Zhang2015}, which may also lead to the discrepancy between the MOP results and measured values.

\begin{figure}[htbp]
  \centering
  \includegraphics[width=0.6\columnwidth]{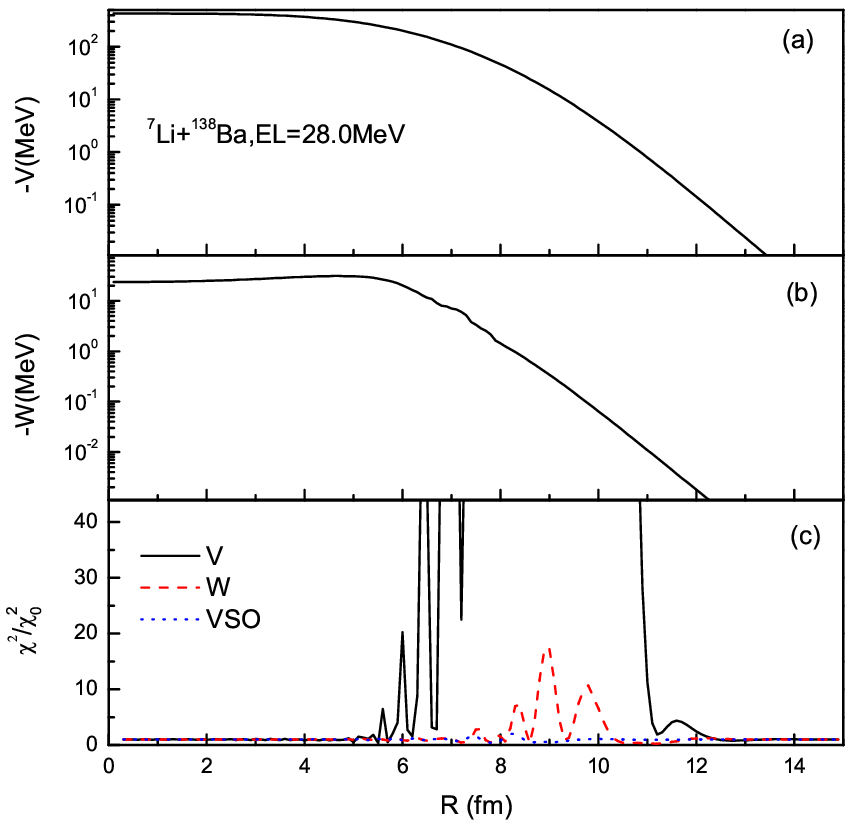}
  \caption{(color online) Notch perturbation analysis of the MOP for the $^7$Li+$^{138}$Ba reaction at EL=28.0 MeV, (a) $V$ of the MOP; (b) $W$ of the MOP; (c) radial sensitivity of the elastic scattering to the MOP. The solid curve, dash curve and dotted curve in (c) represent the results of perturbing $V$, $W$ and $V_{so}$ respectively.}
  \label{f-not-com}
\end{figure}
\begin{figure}[htbp]
  \centering
  \includegraphics[width=0.6\columnwidth]{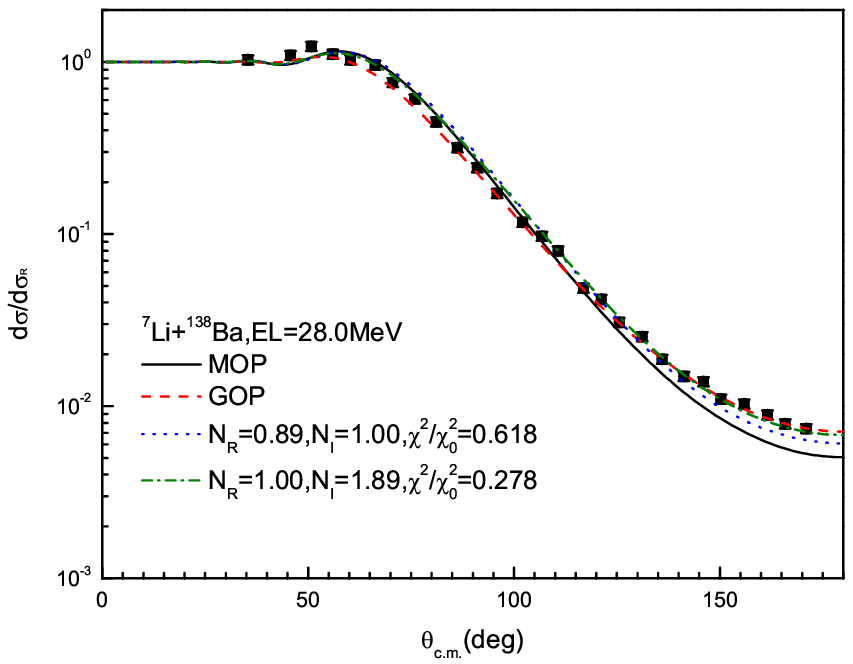}
  \caption{(color online) Calculated elastic-scattering angular distributions in the Rutherford ratio for $^{138}$Ba at EL=28.0 MeV. The solid line, dash line, dotted line, and dash-dotted line denote the results calculated by the MOP, the GOP \cite{Yongli2018}, the MOP with adjusted $V$ and the MOP with adjusted $W$ respectively. The adjustment of $V$ and $W$ are made only in the sensitive region 6 fm$<R<$12 fm.}
  \label{f-not-ex}
\end{figure}
In addition, some discrepancies between the calculated and the measured elastic-scattering angular distributions appear at relatively larger angles, such as the case for $^{138}$Ba target at 28.0 MeV. In order to investigate how to improve the MOP to give a better global agreement with experimental data, notch perturbation method \cite{Moffa1976,Cramer1980} is employed to analyze the sensitivity of the calculated elastic scattering angular distributions to the optical potential. The perturbation is performed by setting $V$, $W$ or $V_{so}$ to 0 in a region of width 0.5 fm centered at radius R. The scattering sensitivity is assessed by ${\chi ^2}/\chi _0^2$, where the ${\chi ^2}$ and $\chi _0^2$ are the chi-squares corresponding to the perturbed and original potentials respectively. $\chi _0^2$ is calculated by
\begin{equation}\label{e-chi2}
\chi _0^2{\text{ = }}\frac{1}{N _{\theta}}\sum\limits_{i = 1}^N {{{\left[ {\frac{{{\sigma ^{T} _0}({\theta _i}) - {\sigma ^{E}}({\theta _i})}}{{\Delta {\sigma ^{E}}({\theta _i})}}} \right]}^2}},
\end{equation}
where ${N _{\theta }}$ is the angle numbers of the experimental elastic-scattering angular distributions for $^7$Li+$^{138}$Ba at EL=28.0MeV. $\sigma ^{T} _0 (\theta _i)$, $\sigma ^{E} (\theta _i)$ and $\Delta \sigma ^{E} (\theta _i)$ represent the theoretical value without perturbation, experimental value and experimental error for the $i$th measured scattering angle respectively. The theoretical value with perturbation, $\sigma ^{T} (\theta _i)$, is used to calculate ${\chi ^2}$ in the same method.

Fig. \ref{f-not-com} shows the MOP for the $^7$Li+$^{138}$Ba system at EL=28.0 MeV and ${\chi ^2}/\chi _0^2$. ${\chi ^2}/\chi _0^2$ for $V_{so}$ of the MOP almost remains at unity, so it is acceptable to ignore the impact from changing $V_{so}$ and focus on only $V$ and $W$. It can be seen that the peaks of ${\chi ^2}/\chi _0^2$ locate mainly in the surface interaction region 6 fm$<R<$12 fm.

We adjust the $V$ and $W$ of the MOP in the sensitive region 6 fm$<R<$12 fm by multiplying $N_{R}$ and $N_{I}$ respectively, and calculate the corresponding ${\chi ^2}/\chi _0^2$. It can be seen in Fig. \ref{f-not-ex} that a better agreement with experimental data at large angles is obtained when $N_{R}$=0.89 and $N_{I}$=1.00. This implies that a weaker real part in the surface region of the MOP may be more suitable for reproducing the measured data. On the other hand, a smaller ${\chi ^2}$ is gotten when $N_{R}$=1.00 and $N_{I}$=1.89, which means that a stronger imaginary part in the surface region may be better. It is expected that the correction of the MOP results from the breakup effect, because the breakup effect, which is not considered in the folding model, just provides a repulsive contribution to the real part and an absorptive contribution to the imaginary part in the surface region.

\section{Summary and conclusion}

A $^7$Li microscopic optical potential without any free parameter is obtained by folding model. The internal wave function of $^7$Li is obtained by the shell model, and a nucleon MOP base on Skyrme nucleon-nucleon effective interaction is adopted. The reaction cross sections and elastic-scattering angular distributions for target from $^{27}$Al to $^{208}$Pb at incident energies below 450 MeV are calculated by the $^7$Li microscopic optical potential. Generally, reasonable agreement with the experimental data is obtained, and the MOP is comparable to the GOP in reproducing the measurements in many cases. However, some discrepancies between the calculated and the measured elastic-scattering angular distributions occur at relatively larger angles. The reason is analyzed, and it is found that the MOP can be improved by adding a repulsive contribution to the real part and an absorptive contribution to the imaginary part in the surface region, which may be achieved by considering the breakup effect. That will be our next subject.

\section*{Acknowledge}
This work is supported by National Natural Science Foundation of China (11705009) and Science Challenge Project (TZ2018005).

\bibliographystyle{unsrt}

\end{document}